\documentclass{ws-ijmpcs}

\begin{document}

\markboth{B. van Soelen, P.J. Meintjes}
{Anisotropic scattering from the circumstellar disc in PSR B1259-63}

\catchline{}{}{}{}{}
%
%%%%%%%%%%%%%%%%%%%%%%%%%%%%%%%%%%%%%%%%%%%%%%%%%%%%%%%%%%%%%%%%%%%%

\title{ANISOTROPIC SCATTERING FROM THE CIRCUMSTELLAR DISC IN PSR B1259-63}

\author{B. VAN SOELEN, P.J. MEINTJES}

\address{Department of Physics, University of the Free State, P.O. Box 339\\Bloemfontein, Free State, 9300, Republic of South Africa\\
VanSoelenB@ufs.ac.za}

\maketitle

\begin{history}
\received{Day Month Year}
\revised{Day Month Year}
\end{history}

\begin{abstract}

The gamma-ray binary system PSR B1259-63 has recently passed through periastron and has been of particular interest as it was observed by Fermi near the December 2010 periastron passage. The system has been detected at very high energies with H.E.S.S.  The most probable production mechanism is inverse Compton scattering between target photons from the optical companion and disc, and relativistic electrons in the pulsar wind. We present results of a full anisotropic inverse Compton scattering model of the system, taking into account the IR excess from the extended circumstellar disc around the optical companion.

\keywords{Radiation mechanisms: non-thermal ; pulsars: individual: PSR B1259−63 ; X-rays: binaries.}
\end{abstract}

\ccode{PACS numbers: 11.25.Hf, 123.1K}

\section{Introduction}

An important Be X-ray Pulsar binary (Be-XPB) example is the TeV gamma-ray system PSR B1259-63, which consists of a 48 ms pulsar orbiting the Be star LS 2883.\cite{johnston92}
The orbit is eccentric ($e \approx 0.87$) and unpulsed TeV gamma-ray emission has been detected by the High Energy Stereoscopic System (H.E.S.S.) close to previous periastron passages.\cite{aharonian05,aharonian07} The circumstellar disc is mis-aligned to
the orbital plane and the pulsar passes through it twice, before and after periastron. Previous observations have also detected unpulsed radio emission near periastron,\cite{johnston99} and X-ray emission across the whole orbit.\cite{chernyakova09}
The system is powered by the spin-down luminosity of the pulsar and the unpulsed radiation is
believed to originate from a stand-off shock-front that forms between the pulsar and stellar winds. The X-ray emission is probably synchrotron radiation and
gamma-rays are produced through inverse Compton (IC) scattering between target photons from the Be star-disc system and relativistic electrons ($ \gamma \sim 10^6$).\cite{tavani97} 
In this study the anisotropic IC scattering in PSR~B1259-63/LS~2883 has been investigated, focusing on the contribution of the infrared (IR) excess from the circumstellar disc to the IC scattering rate. 
The additional effects of photon absorption through pair-production\cite{khangulyan11a} or changes in the shock condition\cite{khangulyan11a,dubus06b} have not been considered here.  This study only presents the effect of the IR excess to investigate whether this a viable source of additional photons and whether this will have a significant influence on the gamma-ray production. 

\section{Modelling}

In order to model the anisotropic IC spectrum from the circumstellar disc in 
PSR~B1259-63 it is necessary to calculate the total number of scatterings, taking into
account the anisotropic scattering rate and the photon contribution from the star and 
the circumstellar disc. The modelling presented here is dependent on the IR excess and curve of growth modelling calculated in Ref.~\refcite{vansoelen11} which assumed the stellar parameters presented by Ref.~\refcite{johnston94}.  For this reason a stellar mass and radius of $10 M_\odot$ and $6 R_\odot$, and a pulsar mass of $1.4 M_\odot$ have been assumed and not the newer system parameters presented by Ref.~\refcite{negueruela11}.

\subsection{Anisotropic Scattering}

The anisotropic scattering rate\cite{aharonian81,fargion97} needs to take into account the scattering angle, $\theta_0$,
between the direction of the observer and the incoming photons over the full solid angle
of the star and disc. The total anisotropic scattering rate is given by\cite{dubus08}
\begin{equation}
 \frac{dN_{\rm tot}}{dt d\epsilon_1} = \int_\Omega \int_\gamma \int_{\epsilon_0} n_{\rm ph}(\epsilon_0) n_{\rm e}(\gamma) \frac{dN_{\gamma,\epsilon_0}}{dt\,d\epsilon_1} \cos \theta d\epsilon_0 d\gamma d\Omega
\end{equation}
where $n_{\rm ph}$ is the number of photons, $n_{\rm e}$ is the electron distribution, $dN/dtd\epsilon_0$ is the anisotropic scattering rate dependent on the scattering angle $\theta_0$, and $\cos\theta$ is the angle to the normal of the photon source (star or disc). The star is assumed to have a black-body
photon distribution ($T_*=25\,000$~K), while the photon distribution from the disc is calculated using the curve of growth method. 

The electron spectrum is assumed to have the form $n_{\rm e}(\gamma) = K_{\rm e} \gamma^{-p}$, where $K_{\rm e} = 1$ is used in this study as only the resulting change due to the IR excess is considered.  The synchrotron radiative cooling time is $t_{\rm sync} \sim 770~ (\gamma/10^6)^{-1}(B/1$~G$)^{-2}$~seconds, which is greater than the light crossing time of the disc. For this reason only a constant electron spectrum has been considered, appropriate for adiabatically cooled electrons.\cite{kirk99}  Therefore, this study considers only whether the IR excess from the large circumstellar disc significantly influences the IC scattering.  Additional effects due to radiative cooling have not been considered here.  

\subsection{Disc photon spectrum - curve of growth method}

The intensity from the circumstellar disc depends on the free-free optical depth through the disc, from which the photon number density is calculated.  The optical depth is determined by the curve of growth method\cite{waters86} and the pulsar-disc orientation. It is assumed that the circumstellar disc has a half-opening angle $\theta_{\rm disc}$, extends to a radius $R_{\rm disc}$ and has a power-law density profile that decreases with distance as $\rho(r)=\rho_0 \left(r/R_*\right)^{-n}$.  

\section{Results}

\begin{figure}[pt]
 \centerline{\psfig{file=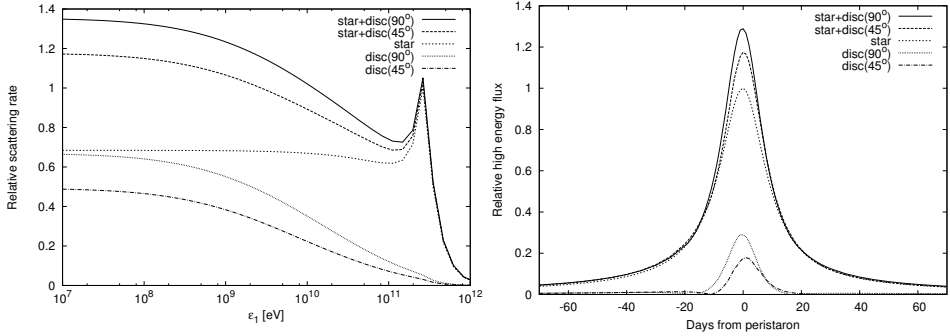,width=12.5cm}}
\vspace*{8pt}
\caption{(Left) Scattering rate for the disc and star contributions in PSR~B1259-63 at periastron for a disc orientated $90^\circ$ and $45^\circ$ to the orbital plane, scaled to the peak stellar contribution.  (Right) The corresponding light curve of the high energy contribution  ($\sim 0.1 - 100$~GeV) of the star and disc around periastron considering only adiabatically cooled electrons.\label{f1}}
\end{figure}

Two orientations of the disc have been
considered; the first with a disc tilted $90^\circ$ to the
orbital plane with the normal lying along the
semi-major axis ($T_{\rm disc}=12\,500$~K \& $\theta_{\rm disc}=5^\circ$), the second titled $45^\circ$ to the
orbital plane with the normal rotated $19^\circ$ from
the semi-major axis ($T_{\rm disc} = 15\,000$~K \& $\theta_{\rm disc} = 0.7^\circ$).\cite{chernyakova06,okazaki11} These disc temperatures and small half-opening angles are consistent with current modelling of the circumstellar discs of Be stars.\cite{okazaki11,telting98}  The disc is assumed to have a radius of $50R_*$, consistent with the binary separation at the time of the binary eclipse.\cite{johnston05,vansoelen11}
The parameters for the curve of growth method were taken from the fit
to previous optical and IR observations\cite{vansoelen11} and
the electron spectrum is modelled using an index of 2.4 between $\gamma = 2.54\times10^5$ and $2.54\times10^7$ (adopted from Ref.~\refcite{kirk99}).
Fig.~\ref{f1} shows the relative scattering rate at periastron scaled to the peak of the stellar contribution for both disc orientations and the corresponding light curve at high energies ($\sim 0.1 - 100$~GeV).
In both cases the contribution of the disc occurs towards high energies (as opposed to very high energies) where it results in an increase in the scattering rate by a factor $\sim 2$.  The light curves show a peak in the disc contribution near periastron and less contribution near the disc crossing where the IC scattering is dominated by the scattering of stellar photons.
The rotation of the disc from the semi-major axis results in a slightly higher flux before periastron and the shifting of the peak away from periastron, while the post-periastron flux is lower because of unfavourable scattering
angles. In both cases the flux increases after the disc crossing as
head-head scattering readily occurs in this region.

\section{Conclusion}

The IC scattering from the circumstellar disc in PSR B1259-63 is shown to increase the flux at high
energies by a factor of $\sim2$ during periastron. Due to the large size of the disc and the increase in the photon density at IR wavelength (increasing by $\sim 1000$ at mid-IR wavelengths)\cite{vansoelen11} it was possible that the
increased solid angle could have a major influence.  However, the flux from the circumstellar disc drops off rapidly with radial distance and the low flux from the disc at the point of the disc crossing mitigates the influence of the disc.  This approximation, considering only adiabatically cooled electrons, shows that the influence of the IR excess does not dramatically increase the scattering rate during the disc crossing. 
In particular, this simplified model does not explain the flare observed by Fermi after the second disc crossing.\cite{abdo11}
It should be noted that a simplified approximation was used to calculate the photon spectrum from
the disc and a more detailed model may result in a higher photon density at the outer edges of the
disc. Additional shock heating effects may also increase the brightness of the circumstellar disc during the pulsar passage. Such an effect could increase the scattering rate near the disc crossing and IC scattering of pre-shock $\gamma=10^4$ electrons may explain the Fermi flare.\cite{khangulyan11}

\section*{Acknowledgments}

BvS is funded by the South African Square Kilometre Array Project.  The numerical calculations made use of the UFS High Performance Computation division. 

% 
% %\begin{thebibliography}{000} %for 3 digits
% %\begin{thebibliography}{00}  %for 2 digits

\end{document}